\documentclass[aip,rsi,preprint,groupedaddress]{revtex4-1}
\usepackage[hypertexnames=false]{hyperref}
\hypersetup{colorlinks=true,citecolor=blue,linkcolor=blue,urlcolor=blue}

\AtBeginDocument{
\renewcommand{\ref}[1]{\autoref{#1}}

}
\usepackage{amsfonts}
\usepackage{graphicx}
\usepackage{lineno}
\usepackage[usenames]{color}

\begin{document}

\title{A broadband Ferromagnetic Resonance dipper probe for magnetic damping
measurements from 4.2 K to 300 K \smallskip{}
}

\author{Shikun He}

\email{skhe@ntu.edu.sg}

\affiliation{Division of Physics and Applied Physics, School of Physical and Mathematical
Sciences, Nanyang Technological University, Singapore 637371}

\author{Christos Panagopoulos}

\email{christos@ntu.edu.sg}

\affiliation{Division of Physics and Applied Physics, School of Physical and Mathematical
Sciences, Nanyang Technological University, Singapore 637371}
\begin{abstract}
A dipper probe for broadband Ferromagnetic Resonance (FMR) operating
from 4.2$\,$K to room temperature is described. The apparatus is
based on a 2-port transmitted microwave signal measurement with a
grounded coplanar waveguide. The waveguide generates a microwave field
and records the sample response. A 3-stage dipper design is adopted
for fast and stable temperature control. The temperature variation
due to FMR is in the milli-Kelvin range at liquid helium temperature.
We also designed a novel FMR probe head with a spring-loaded sample
holder. Improved signal-to-noise ratio and stability compared to a
common FMR head are achieved. Using a superconducting vector magnet
we demonstrate Gilbert damping measurements on two thin film samples
using a vector network analyzer with frequency up to 26$\,$GHz: 1)
A Permalloy film of 5 nm thickness and 2) a CoFeB film of 1.5$\,$nm
thickness. Experiments were performed with the applied magnetic field
parallel and perpendicular to the film plane. 
\end{abstract}
\maketitle

\section{introduction}

In recent years, the switching of a nanomagnet by spin transfer torque
(STT) using a spin polarized current has been realized and intensively
studied.\citep{Slonczewski_STT_1996,Katine_STT_GMR,Mangin_STT_TMR}
This provides avenues to new types of magnetic memory and devices,
reviving the interest on magnetization dynamics in ultrathin films.\citep{STT_block_2014,Brataas_ST_2012}
High frequency techniques play an important role in this research
direction. Among them, Ferromagnetic Resonance (FMR) is a powerful
tool. Most FMR measurements have been performed using commercially
available systems, such as electron paramagnetic resonance (EPR) or
electron spin resonance (ESR).\citep{Farle_review_1998} These techniques
take advantage of the high Q-factor of a microwave cavity, where the
field modulation approach allows for the utilization of a lock-in
amplifier.\citep{Heinrich2006} The high signal-to-noise ratio enables
the measurement of even sub-nanometer thick magnetic films. However,
the operating frequency of a metal cavity is defined by its geometry
and thus is fixed. To determine the damping of magnetization precession,
which is in principle anisotropic, several cavities are required to
study the relation between the linewidth and microwave frequency at
a given magnetization direction.\citep{Heinrich_multi_cavity,Lenz2006,heinrich_2011_YIG_4_cavity}The
apparent disadvantage is that changing cavities can be tedious and
prolong the measurement time. 

Recently, an alternative FMR spectrometer has attracted considerable
attention.\citep{Kalarickal_compareMethod,Harward_2011_70G,Lzu_RSI_2014,Heinrich_broadband_2014,Impact_Conducting_2014,Bailleul_Shield_2013,BilzerMethod07}
The technique is based on a state of the art vector network analyzer
(VNA) and a coplanar waveguide (CPW). Both VNA and CPW can operate
in a wide frequency range hence this technique is also referred to
as broadband FMR or VNA-FMR. The broadband FMR technique offers several
advantages. First, it is rather straightforward to measure FMR over
a wide frequency range. Second, one may fix the applied magnetic field
and acquire spectra with sweeping frequency in a matter of minutes.\citep{BilzerMethod07}
Furthermore, a CPW fabricated on a chip using standard photolithography
enables FMR measurements on patterned films as well as on a single
device.\citep{Neudecker_DeviceFMR_2006} In brief, it is a versatile
tool suitable for the characterization of magnetic anisotropy, investigation
of magnetization dynamics and the study of high frequency response
of materials requiring a fixed field essential to avoid any phase
changes caused by sweeping the applied field.

Although homebuilt VNA-FMRs are designed mainly for room temperature
measurements, a setup with variable temperature capability is of great
interest both for fundamental studies and applications. Denysenkov
et al. designed a probe with variable sample temperature, namely,
4-420$\,$K, \citep{Denysenkov_2003_lowT} however, the spectrometer
only operates in reflection mode. In a more recent effort, Harward
et al. developed a system operating at frequencies up to 70$\,$GHz.\citep{Harward_2011_70G}
However the lower bound temperature of the apparatus is limited to
27$\,$K. Here we present a 2-port broadband FMR apparatus based on
a superconducting magnet. A 3-stage dipper probe has been developed
which allows us to work in the temperature range 4.2$\,$- 300$\,$K.
Taking advantage of a superconducting vector magnet, measurements
can be performed with the magnetic moment saturated either parallel
or perpendicular to the film plane. We also designed a spring-loaded
sample holder for fast and reliable sample mounting, quick temperature
response and improved stability. This setup allows for swift changes
of the FMR probe heads and requires little effort for the measurement
of devices. To demonstrate the capability of this FMR apparatus we
measured the temperature dependence of magnetization dynamics of thin
film samples of Permalloy (Py) and CoFeB in different applied magnetic
field configurations.

\section{apparatus}

\subsection{Cryostat and superconducting magnet system}

\begin{center}
\begin{figure*}
\centering{}\includegraphics[width=0.8\textwidth]{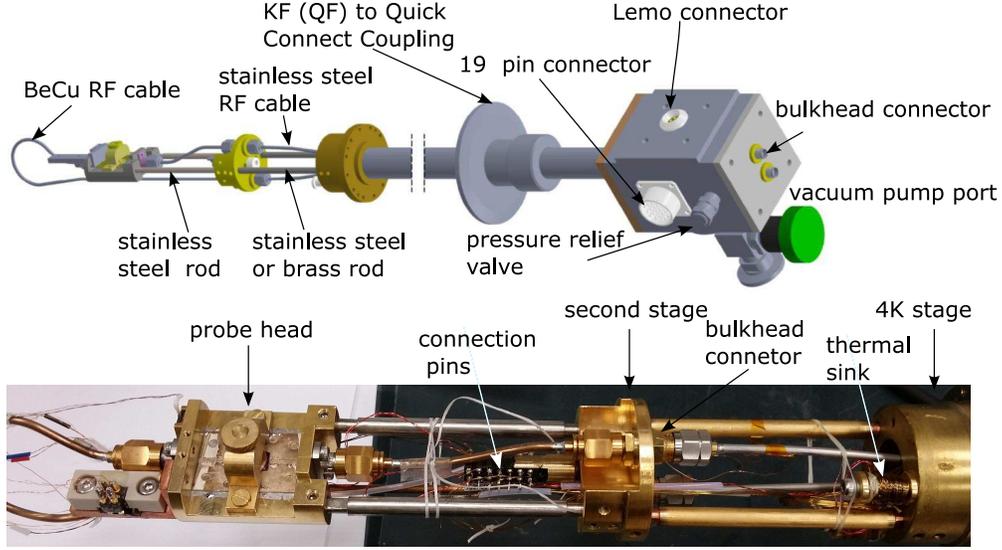}\protect\caption{View of the FMR dipper probe. Top panel: The schematic of the entire
design with a straight type FMR head. All RF connectors are 2.4$\,$mm.
The vacuum cap mounted on the 4$\,$K stage, using In seal, and the
radiation shield mounted on the second stage are not shown for clarity.
Bottom panel: photograph of the components inside the vacuum cap.
\label{fig:Dipper_design}}
\end{figure*}

\par\end{center}

\begin{flushleft}
Our customized cryogenic system was developed by Janis Research Company
Inc. and includes a superconducting vector magnet manufactured by
Cryomagnetics Inc. A vertical field up to 9$\,$T is generated by
a superconducting solenoid. The field homogeneity is $\pm$0.1\% over
a 10$\,$mm region. A horizontal split pair superconducting magnet
provides a field up to 4$\,$T with uniformity $\pm$0.5\% over a
10$\,$mm region. The vector magnet is controlled by a Model 4G-Dual
power supply. Although the power supply gives field readings according
to the initial calibration, to avoid the influence of remnant field
we employ an additional Hall sensor. The cryostat has a 50$\,$mm
vertical bore to accommodate variable temperature inserts and dipper
probes. Our dipper probe described below is configured for this cryostat,
however, the principle can be applied also to other commercially available
superconducting magnets and cryostats.
\par\end{flushleft}

\subsection{Dipper probe}

\ref{fig:Dipper_design} shows a schematic of our dipper probe assembly
and a photograph of the components inside the vacuum cap. The dipper
probe is 1.2$\,$m long and is mounted to the cryogenic system via
a KF50 flange. The sliding seal allows a slow insertion of the dipper
probe directly into the liquid helium space. Supporting arms (not
shown) lock the probe and minimize vibration, with the sample aligned
to the field center. The connector box on top has vacuum tight Lemo
and Amphenol connectors for 18 DC signal feedthrough. Two 2.4$\,$mm
RF connection ports allow for frequencies up to 50$\,$GHz . A vacuum
pump port can be shut by a Swagelok valve. We adopted a three stage
design as shown in the photograph of \ref{fig:Dipper_design}. The
4$\,$K stage and the vacuum cap immersed in the He bath provide cooling
power for the probe. The intermediate second stage acts as an isolator
of heat flow and as thermal sink for the RF cables, providing improved
temperature control. Furthermore, it allows one to change probe heads
conveniently as we discuss later. A separate temperature sensor on
the second stage is used for monitoring purpose. The third stage,
namely, the FMR probe head with the spring loaded sample holder, is
attached to the lower end of the intermediate stage using stainless
steel rods. 

A pair of 0.086'' stainless steel Semi-Rigid RF cables run from the
top of the connector box to the non-magnetic bulkhead connector (KEYCOM
Corp.) mounted on the second stage. BeCu non-magnetic Semi-Rigid cables
(GGB Industries, Inc.) are used for the connection between the second
stage and the probe head. The cables are carefully bent to minimize
losses. The rods connecting the stages are locked by set screws. Loosening
the set screw allows the rod length to be adjusted to match the length
of the RF cables. Reflection coefficient (S$_{11}$) and transmission
coefficient (S$_{21}$) can be recorded simultaneously with this 2-port
design. The leads for the temperature sensors, heater, Hall sensor
and for optional transport measurements are wrapped around Cu heat-sinks
at the 4$\,$K stage before being soldered to the connection pins.

\subsection{Probe head with spring-loaded sample holder}

\begin{figure}
\centering{}\includegraphics[width=8cm]{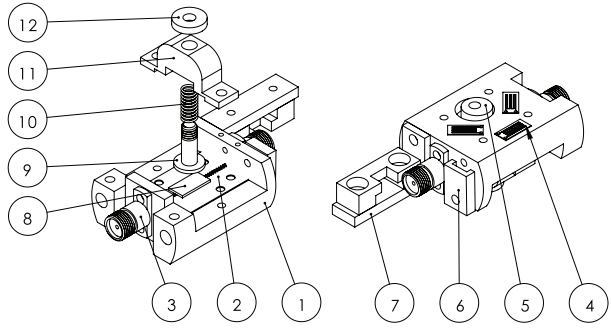}\protect\caption{Schematic of the spring-loaded FMR probe head with straight shape
grounded coplanar waveguide (GCPW). 1 Au plated Cu housing; 2 straight
shape GCPW; 3 flange connector; 4 strain gauge thin film heater; 5
Cernox\textsuperscript{TM} temperature sensor; 6 Hall-sensor housing;
7 housing for 4-pin Dip socket or pingo pin; 8 sample; 9 sample holder;
10 Cu spring; 11 spring housing; 12 sample holder handle nut. \label{fig:SpringLoadeProbeHead}}
\end{figure}

\begin{figure}
\centering{}\includegraphics[width=7.8cm]{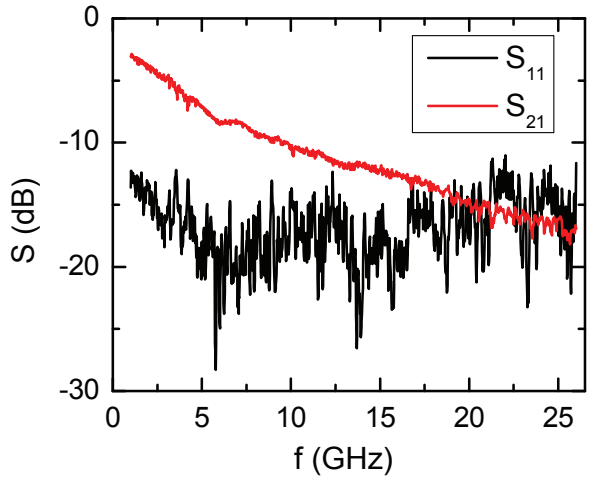}\protect\caption{The reflection (S$_{11}$) and transmission (S$_{21}$) coefficients
of the dipper probe with the straight-line shape GCPW mounted. The
measurement was performed at room temperature. \label{fig:insertion loss} }
\end{figure}

The key part of the dipper probe, namely, the FMR probe head is schematically
shown in \ref{fig:SpringLoadeProbeHead}. The assembly is placed in
a radiation shield tube with an inner diameter of 32$\,$mm. To maximize
thermal conduction between parts, homebuilt components are machined
from Au plated Cu. The 1'' long customized grounded coplanar waveguide
(GCPW) has a nominal impedance of 50 Ohm. The straight-line shape
GCPW was made on duroid\textsuperscript{\textregistered} R6010 (Rogers)
board, with a thickness of 254$\,\mu\text{m}$ and dielectric constant
10.2. The width of the center conductor is 117$\,$$\mu\text{m}$
and the gap between the latter and the ground planes is 76$\,$$\mu\text{m}$.
For the connection, first the GCPW is soldered to the probe head,
and subsequently the center pin of the flange connector (Southwest
Microwave) is soldered to the center conductor of the GCPW. The response
of the dipper with the straight-line shape GCPW installed is shown
in \ref{fig:insertion loss}. The relatively large insertion loss
(-16.9$\,$dB at 26$\,$GHZ) is due to a total cable length of more
than 3$\,$m and multiple connectors. The high frequency current flowing
in the CPW generates a magnetic field of the same frequency. This
RF field drives the precession of the magnetic material placed on
top of the signal line, and gives rise to a change in the system's
impedance, which in turn alters the transmitted and reflected signals.

A spring-loaded sample holder depicted in \ref{fig:SpringLoadeProbeHead}
by items 9 to 12 is designed to mount the sample. The procedure for
loading a sample is as follows: 1) Pull up the handle nut and apply
a thin layer of grease (Apiezon N type) to the sample holder; 2) Place
the sample at the center of the sample holder; 3) Mount the spring-loaded
sample holder to the FMR head; 4) Release the handle nut gradually
so that the spring pushes the sample towards the waveguide. The mounting-hole
of the spring-housing is slightly larger than the outer diameter of
the spring. This allows the sample holder to match the surface of
the GCPW self-adaptively. With the spring-loaded FMR head design,
the sample mounting is simple and leaves no residue from the commonly
used tapes. It maximizes the signal by minimizing the gap between
waveguide and sample, and enhances the stability. Furthermore, it
is suitable for variable temperature measurements due to the enhanced
thermal coupling between the sample, cold head and sensors ( items
9 to 12 in \ref{fig:SpringLoadeProbeHead}.). 

The temperature sensor is mounted at the backside of the probe head.
Due to limited space, the heater consists of three parallel connected
strain gauges with a resistance of 120 Ohm. The Hall sensor can be
mounted according to the required measurement configuration. The position
of the Hall sensor shown in \ref{fig:SpringLoadeProbeHead} is an
example for measurements in the presence of a magnetic field applied
parallel to the sample surface.

\subsection{Probe-head using end-launch connector}

\begin{figure}
\centering{}\includegraphics[width=8cm]{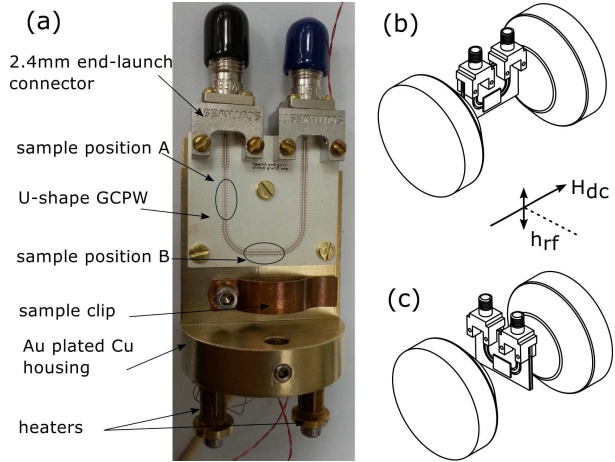}\protect\caption{FMR probe-head with u-shape GCPW and end-launch connector. (a) Photograph
of the probe-head using end-launch connector and U-shape GCPW. Sensors
are mounted at the backside and at the bottom of the Cu housing. Simplified
sketch of the configuration for measuring with an external field generated
by the split coils (b) parallel and (c) perpendicular to the sample
plane. Rotating the dipper probe in the horizontal plane changes from
one configuration to the other. \label{fig:ProbeHead_ELC}}
\end{figure}

Although the probe head with straight-line CPW works well in our experiment,
the necessary replacement of CPW due to unavoidable performance fatigue
over time, or for testing new CPW designs can be time consuming. In
response, end-launch connectors (ELC) utilizing a clamping mechanism
allow for a smooth transition from RF cables to CPW. Soldering the
launch pin to the center conductor of CPW is optional and reduces
the effort for modifications. In \ref{fig:ProbeHead_ELC}, we show
our design of a FMR probe-head using ELC form Southwest Microwave,
Inc. and a homebuilt U-shape GCPW. Similar to the design of \ref{fig:SpringLoadeProbeHead},
a Au plated Cu housing is used to mount the GCPW, ELC and the temperature
and Hall sensors. There are two locations for sample mounting. In
position A, the vertical field is used for measurements with the magnetic
field applied parallel to the surface of the thin film sample whereas
the horizontal field is used for measurements with field perpendicular
to the sample surface. On the other hand, measurements for both configurations
can be accomplished only by using the horizontal field if the sample
is placed in position B. As shown in \ref{fig:ProbeHead_ELC} (b)
and (c), to change between configurations simply requires rotating
the dipper probe by 90 degrees. Nevertheless, we prefer to place the
sample in position A for the parallel configuration since the solenoid
field is more uniform. However, we note that the same design with
the sample placed in position B is suitable also for an electromagnet.
Furthermore, adding a rotary stage to the probe enables angular dependent
FMR measurements. 

\section{Experimental test}

In this section, we present data to assess the performance of the
FMR probe head and discuss two sets of magnetic damping measurements,
demonstrating the capabilities and performance of the appratus.

\subsection{Spring-loaded sample holder}

\begin{figure}
\centering{}\includegraphics[width=8cm]{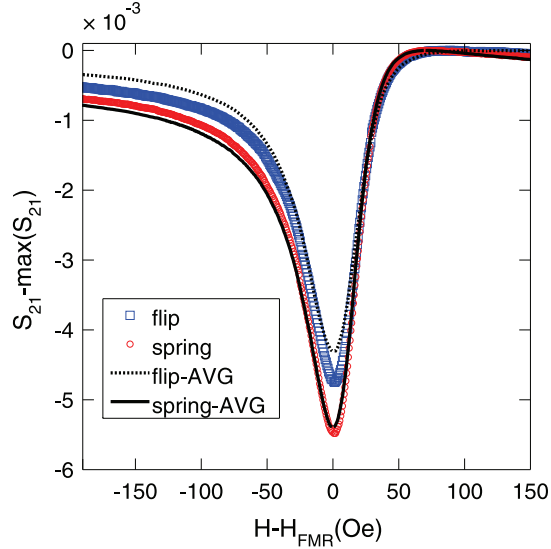}\protect\caption{Comparison between S$_{21}$ signals obtained using spring-loaded
sample holder mounting and flip-sample mounting at 300$\,$K. The
sample has a stack of MgO(3$\,$nm)|CoFeB(2$\,$nm)|MgO(3$\,$nm)
deposited on silicon substrate. (Numbers in parenthesis of the sample
composition represent the thickness of the respective layer.) The
frequency is 10 GHz and the FMR center field is at 1520 Oe.\label{fig:Compar_loadingMethod}}
\end{figure}

We tested our setup using a Keysight PNA N5222A vector network analyzer
with maximum frequency 26.5$\,$GHz. The output power of the VNA is
always 0$\,$dBm in our test. Note that with 2.4$\,$mm connectors
and customized GCPW, our design can in principle operate up to 50$\,$GHz.
The performance of the spring-loaded sample holder is first studied
at room temperature with a 2$\,$nm thick $\textrm{Co}{}_{40}\textrm{Fe}{}_{40}\textrm{B}_{20}$
film. For direct comparison, the FMR spectra are recorded with two
sample loading methods: One with a spring-loaded sample holder (\ref{fig:SpringLoadeProbeHead})
and the other using the common method\citep{Harward_2011_70G}
which only requires Kapton tape. The magnetic field is applied parallel
to the plane of the thin film sample. Six sets of data were obtained
by reloading the sample for each measurement. In \ref{fig:Compar_loadingMethod},
we show the amplitude of the power transmission coefficient from Port
1 to Port 2 (S$_{21}$) at a frequency of 10$\mbox{\,}$GHz and a
temperature of 300$\mbox{\,}$K. The open circles represent a spectrum
for a spring-loaded sample mounting whereas the open squares is the
spectrum showing largest signal for the six flip-sample loadings.
The averaged spectra for all six spectra are shown by solid line and
dotted line, for spring and flip-sample loading, respectively. Two
observations are evident: First, the best signal we obtained using
the flip sample method is approximately 20 percent lower compared
to the spring-loaded method. Thus the spring-loaded method gives a
better signal to noise ratio and sensitivity. Second, for the spring-loaded
method, the difference between the averaged spectrum and single spectra
is negligibly small. On the other hand the variation between measurements
for the flip-sample method can be as large as 20 percent. Hence the
spring-loaded method has better stability and is reproducible.

\subsection{Temperature response}

\begin{figure}
\centering{}\includegraphics[width=7.5cm]{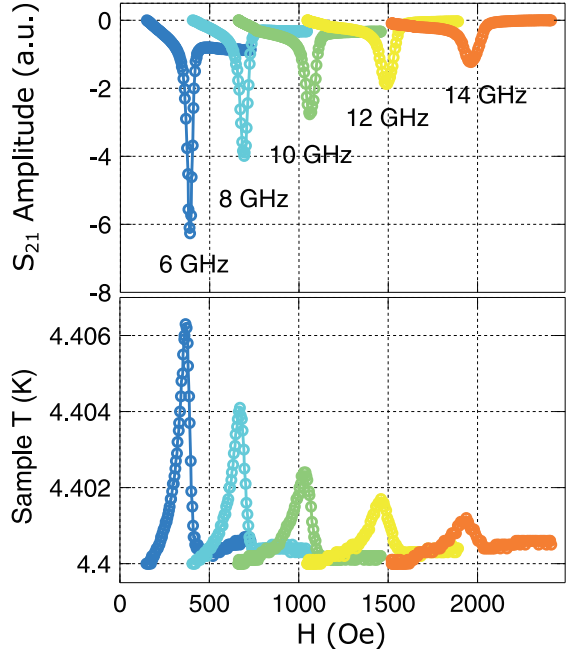}\protect\caption{Sample temperature variation due to FMR at selected frequencies. (upper
panel) Amplitude of S$_{21}$ and (lower panel) temperature variation
of MgO(3$\,$nm)|CoFeB(3$\,$nm)|MgO(3$\,$nm) at 4.4$\,$K measured
with external field parallel to the film plane. \label{fig:Sample-termperature-variation}}
\end{figure}

As detailed in the previous section, the probe head is made of Au
plated Cu blocks with high internal thermal conduction and good thermal
contact. Consequently, the response time of the temperature control
will be small as the characteristic thermal relaxation time of a system
is $C/k$, where $C$ is the heat capacity and $k$ is the overall
thermal conduction. Also, the temperature difference between sample
and sensor is minimized even with the heater turned on. Shown in \ref{fig:Sample-termperature-variation}
are the FMR spectra and temperature variation for a CoFeB film of
3$\,$nm thickness measured at 4.4$\,$K. The external field was swept
at a rate of about -10$\,$Oe/s. For fields close to which FMR peaks
are observed, we detected a temperature rise of a few mK. In fact,
the field values corresponding to maximum temperatures are about 20$\,$Oe
lower than the fields satisfying FMR condition, showing that the characteristic
relaxation time between the sample and cold head is approximately
2 seconds. The temperature rise of the probe head due to FMR indicates
that the magnetic system absorbs energy from the microwave and dissipates
into the thermal bath. Specifically, at the field satisfying the FMR
condition, the damping torque is balanced by the torque generated
by the RF field. However, the dissipation power of such process is
propotional to the thickness of the magnetic film hence is very small.
The successful detection of a temperature rise adds credence to the
high thermal conduction within the probe head and relative low thermal
conduction between different stages. This demonstration shows that
the probe head is capable of measuring samples with phase transitions
in a narrow temperature range, such as a superconducting/ferromagnetic
bilayer system.\citep{Bell_2008} 

\subsection{Magnetic damping measurements}

\begin{figure}
\centering{}\includegraphics[width=8cm]{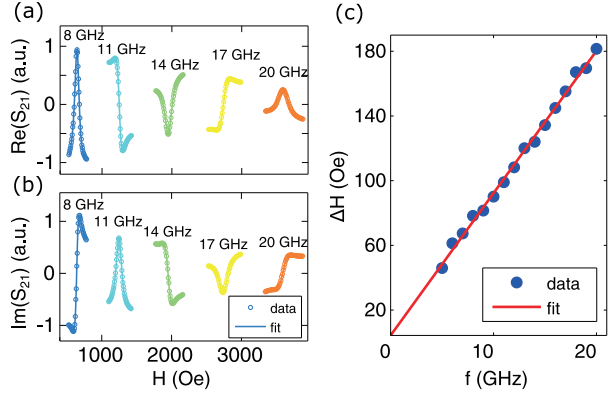}\protect\caption{FMR data of a Py thin film of thickness 5$\,$nm measured at 4.4$\,$K
with magnetic field applied parallel to the sample plane. (a) Real
and (b) Imaginary parts of transmitted signal S$_{21}$ at selected
frequencies. The data are normalized and the relative strength between
the spectra at different frequencies are kept. (c) FMR linewidth as
a function of frequency. The damping was calculated to be 0.012 $\pm0.001$,
using a linear fit. \label{fig:FMR_Py_4K}}
\end{figure}

Although the FMR probe can be used to determine the energy anisotropy
of magnetic materials, our primary purpose is to study magnetic damping
parameter. In the following, two examples of such measurements are
briefly described. Shown in \ref{fig:FMR_Py_4K} is FMR response of
a Py film of 5$\,$nm thickness deposited on a silicon substrate,
measured at 4.4$\,$K. The sweeping external magnetic field is parallel
to the sample surface. Real and Imaginary parts of the spectra obtained
at selected frequencies are plotted with open circles in \ref{fig:FMR_Py_4K}
(a) and (b), respectively. In FMR measurements, the change in the
transmittance, $S_{21}$, is a direct measure of the field-dependent
susceptibility of the magnetic layer. According to the Landau–Lifshitz–Gilbert
formalism, the dynamic susceptibility of the magnetic material in
the configuration where the field is applied parallel to the plane
of the thin film be described as:\citep{BilzerCFB}

\begin{widetext}

\begin{equation}
\chi_{{\rm {\scriptscriptstyle IP}}}=\frac{4\pi M_{{\rm s}}\left(H_{0}+H_{{\rm uni}}+4\pi M_{{\rm eff}}+i\,\Delta H/2\right)}{\left(H_{0}+H_{{\rm uni}}\right)\left(H_{0}+H_{{\rm uni}}+4\pi M_{{\rm eff}}\right)-H_{f}^{2}+i\,(\Delta H/2)\cdot\left[2\left(H_{0}+H_{{\rm uni}}\right)+4\pi M_{{\rm eff}}\right]}\label{eq:SOM_FMR_Chi-IP}
\end{equation}

\end{widetext}

Here, $4\pi M_{{\rm s}}$ is the saturation magnetization, $H_{{\rm uni}}$
is the in-plane uniaxial anisotropy, $4\pi M_{{\rm eff}}$ is the
effective magnetization, $H_{f}=2\pi f/\gamma$, and $\Delta H$ is
the linewidth of the spectrum -- the last term is of key importance
to determine the damping parameter. As shown by solid lines in \ref{fig:FMR_Py_4K}
(a) and (b), the spectra can be fitted very well by adding a background,
a drift proportional to time, and a phase factor.\citep{Nembach2011,Kalarickal_compareMethod}
The field linewidth as a function of frequency -- $\Delta H\,(f)$
is plotted in \ref{fig:FMR_Py_4K} (c). The data points fall on a
straight line. The damping parameter $\alpha_{{\rm {\scriptscriptstyle GL}}}=0.012\pm0.001$
is therefore determined by the slope through\citep{heinrich_inhomo_1991,Lenz2006}:
\begin{equation}
\Delta H=\frac{4\pi}{\gamma}\alpha_{{\rm {\scriptscriptstyle GL}}}\, f+\Delta H_{0}\label{eq:SOM_GilbertDamping}
\end{equation}
The error bar here is calculated from the confidence interval of the
fit.

\begin{figure}
\begin{centering}
\includegraphics[width=8cm]{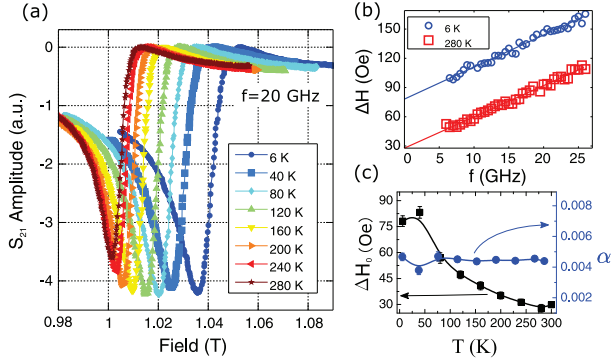}
\par\end{centering}

\protect\caption{Temperature dependent FMR measurement for a CoFeB thin film of thickness
1.5$\,$nm with the magnetic field applied perpendicular to the film
plane. (a) Transmitted FMR signal at 20$\,$GHz obtained at different
temperatures. (b) FMR linewidth as a function of frequency at 6$\,$K
and 280$\,$K. (c) Damping constant and inhomogeneous broadening as
a function of temperature. The solid lines are the guides for the
eye. \label{fig:CFB_T_dependent}}
\end{figure}

We have also tested the setup with a magnetic field applied perpendicular
to the sample plane. The results for a MgO$\,$(3$\,$nm)|Co$_{40}$Fe$_{40}$B$_{20}$(1.5$\,$nm)|MgO$\,$(3nm)
stack deposited on silicon substrate are shown in \ref{fig:CFB_T_dependent}.
Comparing the spectra obtained at different temperatures and fixed
frequency, two observations are evident. First, the FMR peak position
shifts to higher field as the temperature is lowered due to changes
in the effective magnetization. Second, the FMR linewidth increases
with decreasing temperature. Although the interfacial anisotropy can
be determined by fitting the FMR peak positions to the Kittel formula,\citep{kittel1960}
here, we are more interested in the damping parameter as a function
of temperature. The dynamic susceptibility in this configuration is\citep{Devolder2013}: 

\begin{equation}
\chi_{{\scriptscriptstyle {\rm OP}}}=\frac{4\pi M_{{\rm s}}\left(H-4\pi M_{{\rm eff}}-i\,\Delta H/2\right)}{\left(H-4\pi M_{{\rm eff}}\right)^{2}-H_{f}^{2}+i\,\Delta H\cdot\left(H-4\pi M_{{\rm eff}}\right)}\label{eq:SOM_FMR_Chi-OP}
\end{equation}

Following the same procedure as for Py, the real and imaginary part
of the spectra are fitted simultaneously to obtain the linewidth.
In \ref{fig:CFB_T_dependent} (b), we plot the linewidth as a function
of frequency at the two boundaries of our measured temperatures. Although
the measured linewidth at lower temperature is larger, the slope of
the two curves is in good agreement. The additional linewidth at 6$\,$K
is primarily due to zero frequency broadening, which quantifies the
magnitude of dispersion of the effective magnetization. The results
are summarized in \ref{fig:CFB_T_dependent} (c). Gilbert damping
is essentially independent of temperature although there is a minimum
at 40 K. The room temperature value obtained is in agreement with
the value for a thicker CoFeB.\citep{BilzerCFB,Ikeda_2010_Namat}
On the other hand, the inhomogeneous broadening increases with lowering
temperatue. The value at 6$\,$K is more than double compared to room
temperatue. Notably, neglecting the zero-frequency offset $\Delta$H(0),
arising due to inhomogeneity, would give rise to an enhanced effective
damping compared to the intrinsic contribution. Cavity based, angular
dependent FMR may also distinguish the Gilbert damping from inhomogeneity
effects. A shortcoming however, is the need to take into account the
possible contribution of two magnon scattering, which causes increased
complications in the analysis of the data.\citep{Mizukami_PRB-2002,lindnerTMSAngle_2009}
On the other hand, broadband FMR using a dipper probe with the applied
magnetic field in the perpendicular configuration, rules out two magnon
scattering making this technique relatively straightforward to implement.\citep{Arias-PRB-1999} 

The dipper probe discussed here is not limited to measurements of
the damping coefficient. The broadband design is also useful for time-domain
measurements.\citep{timeDomain_PRL_2010} Furthermore, a spin transfer
torque ferromagnetic resonance\citep{STFMR_Liu_PRL_2011} measurement
on a single device can be performed with variable temperature using
bias Tee and a separate sample holder.

\section{conclusion}

We have developed a variable temperature FMR to measure the magnetic
damping parameter in ultra thin films. The 3-stage dipper and FMR
head with a spring-loaded sample holder design have a temperature
stability of milli Kelvin during the FMR measurements. This apparatus
demonstrates improved signal stability compared to traditional flip-sample
mounting. The results for Py and CoFeB thin films show that the FMR
dipper can measure the damping parameter of ultra thin films with:
Field parallel and perpendicular to the film plane in the temperature
range 4.2-300$\,$K and frequency up to at least 26 GHz. 
\begin{acknowledgments}
The authors are grateful to Sze Ter Lim at Data Storage Institute
for preparing the CoFeB samples. We acknowledge Singapore Ministry
of Education (MOE), Academic Research Fund Tier 2 (Reference No: MOE2014-T2-1-050)
and National Research Foundation (NRF) of Singapore, NRF-Investigatorship
(Reference No: NRF-NRFI2015-04) for the funding of this research.
\end{acknowledgments}

\bibliography{RSI-Dipper}

\end{document}